# Nanometre-scale nuclear-spin device for quantum information processing

Y Hirayama[1,2,3], A Miranowicz[2,4,5], T Ota[2], G Yusa[1,2], K Muraki[1], S K Ozdemir[2,5,6] and N Imoto[2,5,6]

[1]NTT Basic Research Labs., NTT Corporation, Atsugi, Kanagawa 243-0198, Japan
[2]SORST Research Team for Interacting Carrier Electronics, JST, Kawaguchi, Saitama 331-0012, Japan
[3]Department of Physics, Tohoku University, Sendai, Miyagi 980-8578, Japan
[4]Institute of Physics, Adam Mickiewicz University, 61-614 Poznan, Poland
[5]Graduate School of Engineering Science, Osaka University, Toyonaka, Osaka 560-8531, Japan
[6]CREST Photonic Quantum Information Project, JST, Kawaguchi, Saitama 331-0012, Japan

E-mail: hirayama@nttbrl.jp, Phone: 046-240-3330, Fax: 046-240-4317

**Abstract** We have developed semiconductor point contact devices in which nuclear spins in a nanoscale region are coherently controlled by all-electrical methods. Different from standard nuclear-magnetic resonance technique, the longitudinal magnetization of nuclear spins is directly detected by measuring resistance, resulting in ultra-sensitive detection of the microscopic quantity of nuclear spins. All possible coherent oscillations have been successfully demonstrated between two levels from four nuclear spin states of $I$=3/2 nuclei. Quantum information processing is discussed based on two fictitious qubits of an $I$=3/2 system and methods are described for performing arbitrary logical gates both on one and two qubits. A scheme for quantum state tomography based on $M_z$-detection is also proposed. As the starting point of quantum manipulations, we have experimentally prepared the effective pure states for the $I$=3/2 nuclear spin system.

## 1. Introduction

The coherent control of quantum states, e.g., qubit operation, in solid-state systems has attracted much interest from the viewpoint of developing a scalable quantum computer [1-3]. Among the many candidates for solid-state qubits, quantum nanostructures based on semiconductor systems have ideal characteristics for confining a small number of quantities to coherently control exciton, charge, spin or nuclear spin. All-optical control has been demonstrated for exciton confined in a semiconductor quantum dot [4-6]. Electron charge [7,8] and/or spin [9] has been coherently manipulated in a coupled quantum dot, resulting in all-electrical qubit operation.

On the other hand, nuclear spins have well-defined coherent characteristics as evidenced by nuclear magnetic resonance (NMR) and magnetic resonance imaging (MRI) techniques. Quantum information processing using NMR of spins 1/2 nuclei has still been the forerunner for experimental quantum computation since the pioneering works of Cory et al. [10] and Gershenfeld and Chuang [11] (for reviews, see [12-14] and references therein). Although there may be a problem with future scalability, 7-qubit operation has been demonstrated using a special macromolecule [15]. Therefore, it is naturally exciting to extend such coherent operation of nuclear spins in solid-state systems. An interesting nuclear spin qubit has been proposed by assuming the manipulation of individual nuclear spins [16, 17], and some experimental trials have attempted to arrange nuclear spin, i.e. atom, in semiconductors [18]. However, we will need more time and efforts to achieve this ultimate goal.

In experiments, manipulation of nuclear spins has been achieved for a group of nuclear spins in semiconductor systems. For example, nuclear spins have been polarized using an optical method based on illumination with circularly polarized light [19, 20]. The circular polarization results in the generation of spin-polarized electrons, and nuclear spins are polarized via spin flip-flop process when the spin polarized electrons relax to a spin-unpolarized equilibrium state. Nuclear spin polarization in semiconductors can also be achieved by using electrical means. Examples are nuclear spin polarization at the edge of a quantum Hall bar [21, 22] and in a quantum dot in a spin-blockade condition [23]. Electrical polarization of nuclear spins has also been studied on the fractional quantum Hall system [24-26]. All of these approaches use a special situation where different electron spin states degenerate each other. This degeneracy enables us to have flip-flop interaction between electron and nuclear spins while keeping energy and momentum conservation rules.

In our experiments, we use the fractional quantum Hall regime at a Landau-level filling factor of $\nu = 2/3$, in which coupling of nuclear spins to conduction electrons is known to be pronounced [24-26]. Electron-nuclear spin interactions at the degenerate points between spin polarized and unpolarized $\nu=2/3$ states are evidenced by a gradual increase in resistance, i.e. the $R_{xx}$ value, when the sample is driven by a certain current flow [24]. The gradual resistance enhancement reflects the gradual polarization of nuclear spins. The unique feature of this interaction in the $\nu = 2/3$ degenerate situation is that the $R_{xx}$ value is approximately proportional to the total longitudinal magnetization, $M_z$, coming from nuclear spin polarization [25]. Although the mechanism of this proportionality is not clear yet, this simple relation is helpful in realizing an ultra-high sensitive detection of nuclear spin polarization based on resistively detected NMR [27].

The constituent atoms, Ga and As, in our GaAs device have an $I=3/2$ nuclear spin system and the states separate into four. An interesting generalization of the original approach to using nuclear spin-3/2 for quantum information processing was first suggested by Kessel and Ermakov [28] and partially realized by Khitrin and Fung [29]. Since then, the NMR quantum information methods for spin-3/2 systems have further been developed [30-39] in close analogy to standard NMR quantum information processing with spin-1/2 systems. The vast majority of the experiments on NMR quantum information processing have been performed in liquids, liquid crystals, or powders under magic angle spinning conditions. Only a few NMR quantum information processing experiments with single-crystal solids have been reported [12, 32, 37]. However, most experiments have been based on the conventional NMR technique, which reads out $M_{xy}$, although Leuenberger et al. [38, 39] have taken into account nuclear spin control in semiconductors.

In this paper, we describe a nanometre-scale spin-3/2 device in which nuclear spin polarization is detected by the resistance value with ultra-high sensitivity. The coherent control between any two states from four spin levels has been demonstrated for the microscopic quantity of nuclear spins confined in the GaAs point-contact nanometre regime [27]. The experimental results clearly demonstrate arbitrary superposition among four spin states, i.e., *quartit* operation, which is equivalent to two-qubit operation. Here, |-3/2>, |-1/2>, |1/2>, and |3/2> of As (or Ga) nuclear spin states can be considered as |11>, |10>, |01>, and |00> states of the fictitious two-qubit system. In addition, we will show the pulse operations necessary for implementing rotations, Hadamard, and controlled-not (CNOT) gates for the quartit system. State tomography, which requires the readout of all matrix elements in the density matrix after the quantum operations, will be discussed in connection with an exchange of matrix elements and $M_z$-detection. It is also important to control the populations among the four spin states as the initialization before quantum operations. We will describe an experimental demonstration of the preparation of effective pure states in our device in the last part of this article.

## 2. Coherent control of nuclear spins in a semiconductor nanodevice
*2.1. Point-contact device for nuclear spin manipulation*
We fabricated a monolithic semiconductor device integrated with a point contact channel and an antenna gate for the coherent control of nuclear spins in nanometre scale (see figure 1). The point contact is a narrow constriction of a two-dimensional electron gas defined by a depletion region under the split Schottky gates. By controlling the backgate bias and vertical static magnetic field $B_0$, the point contact region was set to a $v$ =2/3 degenerate situation. This degeneracy was confirmed using a diagram of $R_{xx}$ as a function of $B_0$ and electron density [40]. The spin unpolarized and polarized $v$ =2/3 states faced each other through a transition region with a high $R_{xx}$ value. In this transition region, the degeneracy was realized and a strong interaction was achieved between electron and nuclear spins as evidenced by a gradual $R_{xx}$ enhancement as shown in figure 2. The gradual resistance enhancement reflects nuclear spin polarization by electron-spin-flipping through the contact hyperfine interaction when sufficient current is driven through the system (dynamic nuclear spin polarization). The current density becomes very high in the constricted region so that nuclear spin polarization occurs only in the point contact region. The experimental result in figure 2 where no resistance enhancement appears without depletion under the Schottky gates strongly supports a selective polarization of nuclear spins in the point contact region.

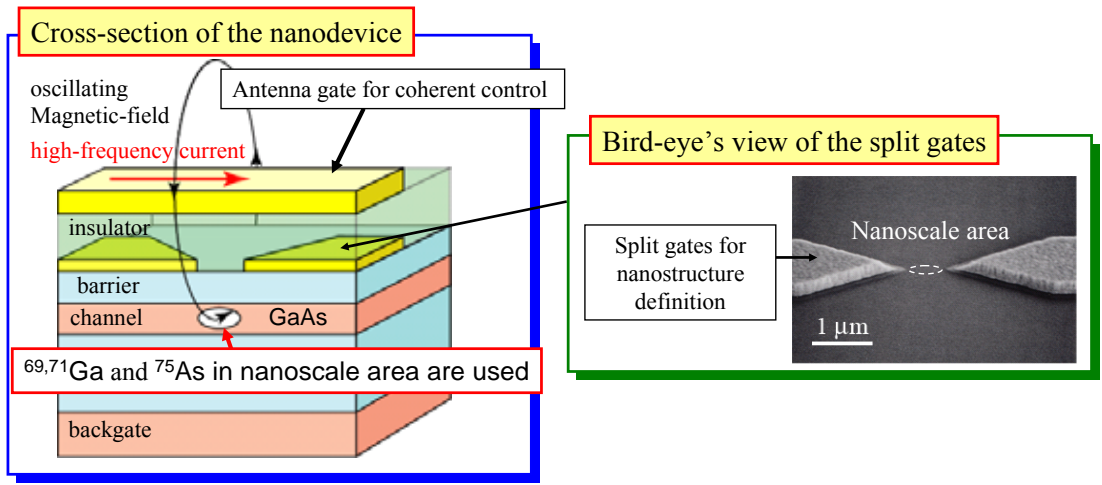

**Figure 1** Schematic diagram illustrating the main aspects of the semiconductor device used for coherently controlling nuclear spins in a nanometre-scale region. The structure contains a 20-nm GaAs quantum well with AlGaAs barrier layers grown on n-GaAs (100) substrate that functions as a back-gate to control electron density. A pair of Schottky gates defines the point contact channel, which is indicated by a white ellipse. The gap between the split Schottky gates is 600 nm. The static magnetic field $\mathbf{B_0}$ is applied perpendicular to the grown surface.

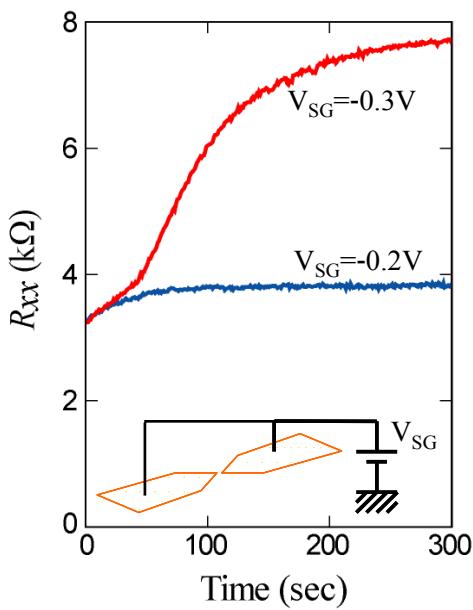

**Figure 2** $R_{xx}$ enhancement from the dynamic nuclear spin polarization near the transition point of $v$=2/3. Two-dimensional electrons under the split Schottky gates are not completely depleted at $V_{SG}$ =-0.2 V and $R_{xx}$ enhancement is weak in this case. On the other hand, strong $R_{xx}$ enhancement is observed for $V_{SG}$ =-0.3 V, where electrons under the split gates are completely depleted and the point contact channel is well defined. The results clearly support the occurrence of the local nuclear spin polarization in the point contact region.

Furthermore, a gradual increase in the resistance means that the polarization of nuclear spins in the point contact region can be measured from the resistance between both ends of the point contact. Actually, the resistance value is approximately proportional to the total longitudinal magnetization of nuclear spin $M_z$ in our experiments. This is well verified by a comparison between experimentally observed coherent oscillations, discussed in the following section, and simulation based on the assumption of $\Delta M_z \propto \Delta R_{xx}$ [27].

The antenna gate was integrated in our device as shown in figure 1. Radio-frequency (r.f.) current through this antenna resulted in local irradiation of the point contact with r.f. field. The nuclear spins in the point contact region were coherently manipulated by this r.f. field.

*2.2. Coherent oscillation among four spin levels*

The point contact is defined in the GaAs quantum well in our sample and consists of $^{69,71}$Ga and $^{75}$As each having total spin $I$=3/2. Thus, for each nuclide under static magnetic field $B_0$, four equally spaced energy states with energy separation of $\hbar\omega_0$ are formed by the Zeeman effect. The additional electric quadrupolar interaction shifts by energies $\Delta_q$ and $-\Delta_q$ for $|\pm 3/2\rangle$ and $|\pm 1/2\rangle$ states, respectively. The energy difference between adjacent states is, therefore, shifted by $2\Delta_q$ from $\hbar\omega_0$, allowing three possible resonances at $\hbar\omega_0 - 2\Delta_q$, $\hbar\omega_0$, and $\hbar\omega_0 + 2\Delta_q$ as shown in figure 3(a).

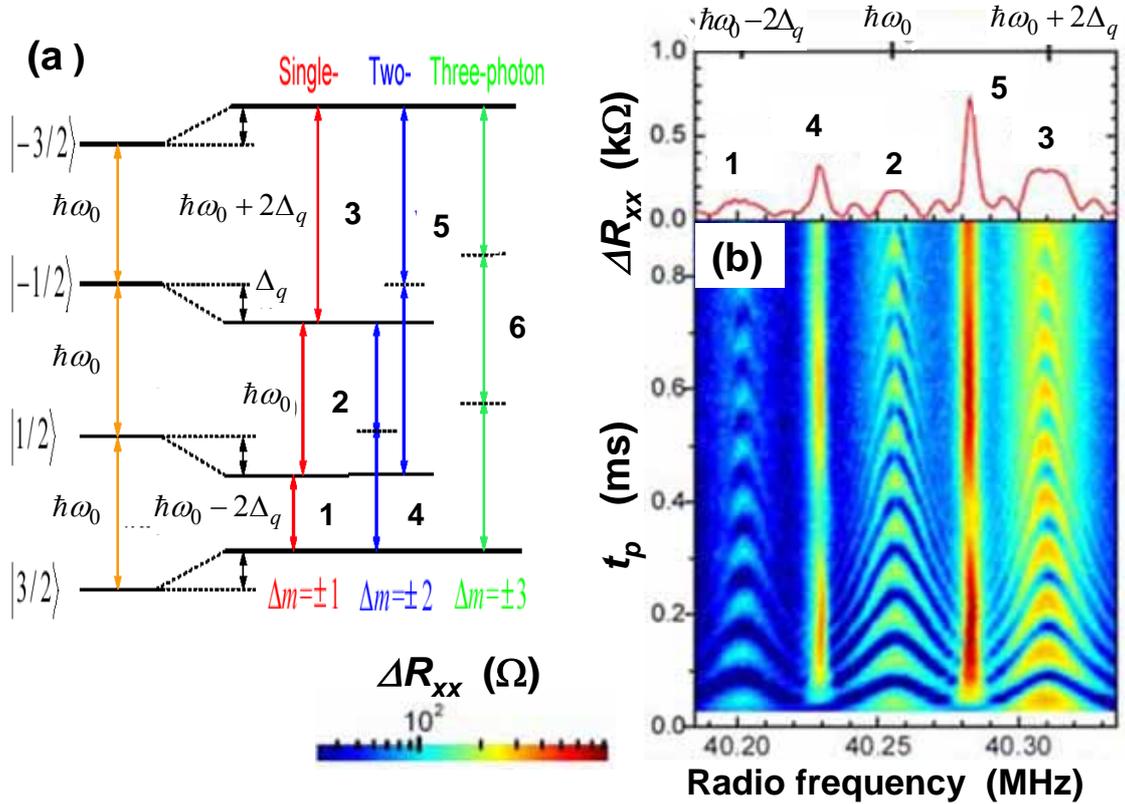

**Figure 3** (a) Schematic energy level diagram of nuclear spin states for $I$=3/2 with (right) and without (left) electric quadrupolar interactions. (b) Coherent oscillations originating from various transitions between the four spin states observed near the NMR resonance frequency of $^{75}$As. The measurement was performed at $B_0$ = 5.5 T at 0.1 K. $\Delta R_{xx}$ is color-plotted. In addition to the single quantum transitions (1, 2 and 3), two double quantum transitions (4 and 5) are clearly visible. A triple quantum transition (6) was also observed when the strength of the radiation was increased by driving greater alternating current through the antenna gate (data not shown). The upper column shows the NMR spectrum measured at $t_p$ =0.126 ms.

After saturation of the $R_{xx}$ value, or in other words nuclear spin polarization as the intialization, pulsed r.f. current is applied to the antenna gate (see figure 1). This pulse current produces pulsed r.f. magnetic field $B_1$ which is proportional to a square root of the input power to the antenna gate. When the applied radio frequency is in resonance with NMR frequency, for example $\omega_0 - 2\Delta_q/\hbar$, $\omega_0$, or $\omega_0 + 2\Delta_q/\hbar$, the superposition coherently rotates between two states, resulting in the oscillatory change in longitudinal magnetization $M_z$. These oscillations are finally detected by resistance change $\Delta R_{xx}$ as shown in figure 3(b). Here, we define $\Delta R_{xx}$ as positive when $R_{xx}$ drops after the r.f. pulse. The resistance measurement corresponds to the projection of the probability amplitude of the nuclear spin states in our experiments. The oscillation at $\hbar\omega_0$ corresponds to the rotation whose path describes a great circle passing through localized states $|1/2\rangle$ and $|-1/2\rangle$ in a geometrical representation of superposition states of a two-level system (Bloch sphere). Similar clear oscillations were observed for the states between $|3/2\rangle$ and $|1/2\rangle$ and between $|-1/2\rangle$ and $|-3/2\rangle$, corresponding to transitions 1 and 3 in figures 3(a) and (b).

According to energy and angular momentum conservation rules, several transitions are possible in addition to a single quantum coherence between levels separated by a single quantum of angular momentum ($\Delta m = 1$). Two double quantum coherent oscillations appear in the four-state system with $I$=3/2 between two levels separated by two quanta of angular momentum ($\Delta m = 2$) driven by two-photon transitions. Furthermore, triple quantum coherent oscillation is possible between two levels separated by $\Delta m = 3$ driven by three-photon transitions as shown in figure 3(a). Actually, we can observe additional oscillatory features [4 and 5 in figure 3(b)] between three single quantum oscillations, reflecting double quantum coherent oscillations, which have resonance r.f. frequency of $\omega_0 - \Delta_q/\hbar$ and $\omega_0 + \Delta_q/\hbar$.

The period of coherent rotations decreases with increasing $B_1$ and the multiple (double and triple) quantum coherent oscillations become more visible. The triple quantum coherent oscillation was observed at the frequency of $\hbar\omega_0$ under the large $B_1$ because the energy difference between $|3/2\rangle$ and $|-3/2\rangle$ is equal to $3\hbar\omega_0$ regardless of the quadrupolar interaction [figure 3(a)]. Furthermore, most of the observed features are well explained by numerical simulation using the rotation frame approximation [39, 41], where a linear conversion efficiency is assumed between $\Delta M_z$ and $\Delta R_{xx}$. The simulations can reproduce not only the periods of all observed oscillations but also the behaviors of off-resonance tails. Their widths for higher $\Delta m$ are scaled down by the photon number involved and are therefore narrower [27, 41]. Good agreement between simulations and experiments confirm the validity of our detection scheme through $\Delta M_z \propto \Delta R_{xx}$, although the mechanism behind this linearity is not clear yet.

Similar coherent oscillations were clearly observed for $^{69}$Ga and $^{71}$Ga (data not shown) as well as for $^{75}$As. This means that three single, two double and one triple quantum coherences for one nuclide, 18 in total for the three nuclides, are completely controlled by all-electrical means. The results in figure 3(b) confirm arbitrary control of superposition between four spin states, $|3/2\rangle$, $|1/2\rangle$, $|-1/2\rangle$ and $|-3/2\rangle$, and this is equivalent to two-qubit operation. We will discuss basic quantum operations for $I$=3/2 nuclear spin system in section 3.

Apart from the coherent control, the obtained results are interesting as a novel NMR technique. Based on an estimated volume of the nanoscale point contact region of approximately 200×200 nm (area) and 10 nm (thickness), the number of nuclear spins involved is of the order of ~$10^8$ or less, which is much less than the detection limit of conventional NMR ($10^{11}$-$10^{13}$). The clear NMR signal indicates that very highly sensitive NMR has been achieved in the point contact device [27, 42]. Furthermore, $\Delta M_z$ is directly detected by $\Delta R_{xx}$ so that this NMR is suitable for the systems with $I$>1/2, which is again difficult to effectively detect with the conventional NMR technique. Although the degeneracy of the different electron spin states at $\nu$=2/3 is a special situation, the obtained result suggests the possibility of realizing extremely highly sensitive NMR on a chip by taking advantage of the coupling between nuclear spins and conduction electrons in solid-state systems.

*2.3. Decoherence in nanoscale NMR device*

Decoherence is the most important issue for a solid-state qubit, although nuclear spins are generally expected to have a long decoherence time $T_2$. Here, we estimated $T_2$ by fitting the single quantum coherent oscillation between $|-1/2\rangle$ and $|-3/2\rangle$ of $^{75}$As with a damped oscillation, $\propto 1-\cos(\Omega_R t_p)\exp(-t_p/T_2)$, where $\Omega_R$ is the frequency of the coherent oscillation and $t_p$ is the r.f. pulse duration. A rather small $B_1$ was used to avoid additional effects from the double quantum oscillations. Figure 4(a) shows the oscillation without any decoupling process. $T_2$ becomes around 0.6 ms. Our device allows us to decouple nuclei from the electron, since electrons in the point-contact region can be depleted during the manipulation of nuclei with the r.f. pulse. This is achieved by applying a larger negative voltage to the split gates. The resonance frequency shifts about 9 KHz due to the Knight shift [43]. Figure 4(b) shows the coherent oscillation obtained with this electron-nuclear spin decoupling. $T_2$ is found to be enhanced to 1.5 ms.

We extracted decoherence rates, $1/T_2$, to which individual mechanisms contribute, from the set of decoupling experiments [43]. The obtained results reveal that electron mediated As-Ga coupling and direct dipole coupling of As-As play an important role in the dephasing. On the other hand, the heteronuclear direct dipole coupling (As-Ga) only makes a small contribution to the decoherence, which seems counterintuitive because As and Ga are located at the nearest neighbor each other (with distance of 0.433$a$, where $a$ =0.565 nm is the lattice constant of GaAs ). This feature may be explained by a zinc-blend lattice structure of GaAs crystal. The dipole coupling is proportional to $|\mathbf{r}|^{-3}(3\cos^2\theta-1)$, where $\mathbf{r}$ is the vector between two nuclei and $\theta$ represents the angle between $\mathbf{r}$ and vector $\mathbf{B_0}$ [44]. Our structure was fabricated from a heterostructure grown on (100) surface and the $\mathbf{B_0}$ was perpendicular to the layer structures so that, for all nearest As and Ga bonds, $\theta$ becomes 54.7347 deg, which satisfies $3\cos^2\theta-1=0$ (i.e.,the magic angle). When the conductive electrons are not depleted, the As-Ga coupling makes a strong contribution because As and Ga nuclei can couple with the shortest distance via conduction electrons. The obtained results support the possibility of manipulating the coupling between Ga and As by controlling the electron density via gate bias. We may be able to connect nuclear spins separated in different positions with conductive electrons for future qubit devices based on nuclear spins.

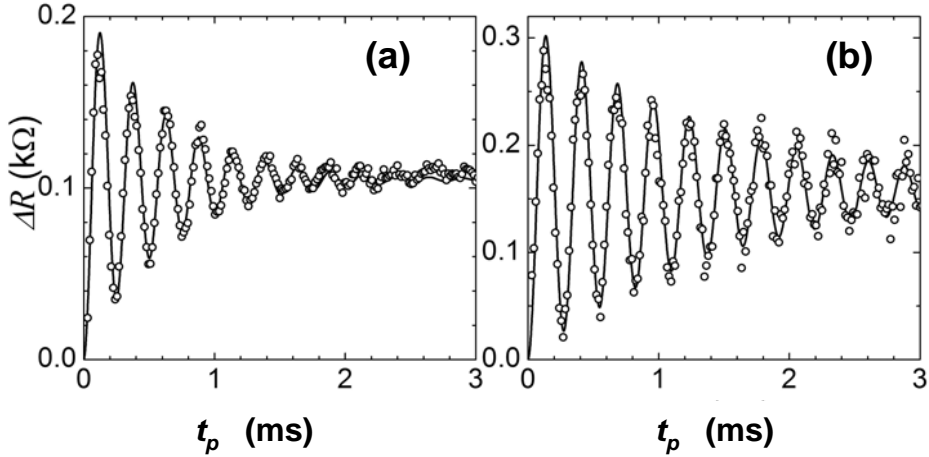

**Figure 4** Coherent oscillations between $|-1/2\rangle$ and $|-3/2\rangle$ with and without electron-nuclear spin decoupling. (a) Without any decoupling. (b) With electron-nuclear spin decoupling by the depletion of the point contact region with more negative $V_{SG}$.

## 3. Quantum gate operations for nanoscale NMR device
### 3.1. Quantum information processing with a quartit

A quadrupolar spin-3/2 nuclei is a four-level system, which is known as a *quartit* in the quantum information context and formally equivalent to two qubits. This can be seen by identifying all the quartit states with the states of two logical (fictitious) qubits (say *A* and *B*), as in the following examples:

$$|0\rangle \equiv |00\rangle \equiv |0\rangle_A |0\rangle_B \equiv |3/2\rangle,$$
$$|1\rangle \equiv |01\rangle \equiv |0\rangle_A |1\rangle_B \equiv |1/2\rangle,$$
$$|2\rangle \equiv |10\rangle \equiv |1\rangle_A |0\rangle_B \equiv |-1/2\rangle,$$
$$|3\rangle \equiv |11\rangle \equiv |1\rangle_A |1\rangle_B \equiv |-3/2\rangle. \quad (1)$$

Thus, any two-qubit input state can be written as a quartit state:

$$|\psi\rangle = c_0 |0\rangle + c_1 |1\rangle + c_2 |2\rangle + c_3 |3\rangle \equiv [c_0, c_1, c_2, c_3]^T, \quad (2)$$

where the complex amplitudes $c_i$ are normalized to unity.

One of the fundamental theorems of quantum information says that a set of gates composed of single qubit rotations and CNOT is universal for quantum computation, i.e., arbitrary unitary operations can be performed on any number of qubits [45]. However, the question is whether these gates can be realized not on real qubits but on fictitious qubits in a quartit using the NMR schemes, in particular, whether one can selectively rotate logical qubit *A* without rotating logical qubit *B* and vice versa.

Here, we describe methods for performing logic gates using nuclei with spin 3/2 applicable for our solid-state device [27]. Various two-qubit methods for quantum state engineering and quantum information processing have already been tested experimentally in spin-3/2 NMR systems, including demonstrations of (a) classical [29, 30] and quantum [32, 33, 37] logical gates, (b) the generation of Bell states [32, 33], (c) Grover's search algorithm [31, 37, 39], (d) quantum state [35] and process [37] tomography schemes, (f) the Deutsch-Jozsa algorithm [34, 37], and (g) quantum Fourier transformation [37]. It should be stressed that many of the above algorithms have been realized only in liquid-state NMR or have only been partially experimentally tested.

### 3.2. Single logical qubit gates in a quartit

Any single-qubit (spin-1/2) unitary gate can be described by

$$\hat{U} = e^{i\alpha} \hat{R}_\mathbf{n}(\theta), \quad (3)$$

where $\hat{R}_\mathbf{n}(\theta)$ is a rotation in the Bloch sphere over an angle $\theta$ about the axis $\mathbf{n} = (n_x, n_y, n_z)$ and can be given by

$$\hat{R}_\mathbf{n}(\theta) \equiv \exp(-\frac{i\theta}{2} \mathbf{n} \cdot \hat{\boldsymbol{\sigma}}) = \hat{\sigma}_I \cos\frac{\theta}{2} - i(n_x \hat{\sigma}_x + n_y \hat{\sigma}_y + n_z \hat{\sigma}_z) \sin\frac{\theta}{2}, \quad (4)$$

in terms of the Pauli matrices $\hat{\boldsymbol{\sigma}} = (\hat{\sigma}_x, \hat{\sigma}_y, \hat{\sigma}_z)$ and identity operator $\hat{\sigma}_I$. In special cases, rotations about the axes $k = x, y, z$ are simply described by

$$\hat{X}(\theta) = \begin{bmatrix} \cos\frac{\theta}{2} & -i\sin\frac{\theta}{2} \\ -i\sin\frac{\theta}{2} & \cos\frac{\theta}{2} \end{bmatrix}, \hat{Y}(\theta) = \begin{bmatrix} \cos\frac{\theta}{2} & -\sin\frac{\theta}{2} \\ \sin\frac{\theta}{2} & \cos\frac{\theta}{2} \end{bmatrix}, \hat{Z}(\theta) = \begin{bmatrix} e^{-i\theta/2} & 0 \\ 0 & e^{i\theta/2} \end{bmatrix}, \quad (5)$$

where $\hat{R}_{x,y,z}(\theta)$ are denoted by $\hat{X}(\theta)$, $\hat{Y}(\theta)$, $\hat{Z}(\theta)$, respectively. Single real qubit rotations can be directly achieved by NMR techniques (see, e.g., [41] or any review on NMR quantum computation, e.g., [14]).

Selective rotations in a quartit can be realized by applying transition selective r.f. pulses at the resonant frequency between two energy levels, say $|m\rangle$ and $|n\rangle$. The relations in (5) can be readily generalized to $4\times 4$ matrices $\hat{X}_{nm}(\theta)$, $\hat{Y}_{nm}(\theta)$, and $\hat{Z}_{nm}(\theta)$ describing the corresponding rotations between levels $|m\rangle$ and $|n\rangle$, while leaving the other levels unchanged.

Rotations of logical qubit $B$ in a quartit over an angle $\theta$ about axes $x,y,z$ can be realized by applying r.f. $\theta$-pulses at frequencies $\omega_0 - 2\Delta_q / \hbar$ and $\omega_0 + 2\Delta_q / \hbar$, i.e.:

$$\hat{X}^B(\theta) = \hat{X}_{01}(\theta)\hat{X}_{23}(\theta),$$
$$\hat{Y}^B(\theta) = \hat{Y}_{01}(\theta)\hat{Y}_{23}(\theta),$$
$$\hat{Z}^B(\theta) = \hat{Z}_{01}(\theta)\hat{Z}_{23}(\theta). \tag{6}$$

Here, we use a notation where the pulses are applied from the right to the left, which is the opposite order to that represented in standard graphical schemes. Note that the pulses can be applied simultaneously or sequentially in arbitrary order if the corresponding operators commute [like in (6)]. Analogously, rotations of logical qubit $A$ in a quartit can be obtained by applying r.f. pulsing at frequencies $\omega_0 - \Delta_q / \hbar$ and $\omega_0 + \Delta_q / \hbar$ to induce the two-photon transitions (double-quantum transitions) as follows:

$$\hat{X}^A(\theta) = \hat{X}_{02}(\theta)\hat{X}_{13}(\theta),$$
$$\hat{Y}^A(\theta) = \hat{Y}_{02}(\theta)\hat{Y}_{13}(\theta),$$
$$\hat{Z}^A(\theta) = \hat{Z}_{02}(\theta)\hat{Z}_{13}(\theta). \tag{7}$$

Rotations of logical qubit $A$ can also be obtained by applying r.f. pulses at frequencies $\omega_0 - 2\Delta_q / \hbar$, $\omega_0$ and $\omega_0 + 2\Delta_q / \hbar$ (single-quantum transitions) by using the relations discussed in the Appendix:

$$\hat{X}^A(\theta) = \hat{Y}_{12}(\pi)\hat{X}_{01}(\theta)\hat{X}_{23}(-\theta)\hat{Y}_{12}(-\pi),$$
$$\hat{Y}^A(\theta) = \hat{Y}_{12}(\pi)\hat{Y}_{01}(\theta)\hat{Y}_{23}(-\theta)\hat{Y}_{12}(-\pi),$$
$$\hat{Z}^A(\theta) = \hat{Y}_{12}(\pi)\hat{Z}_{01}(\theta)\hat{Z}_{23}(\theta)\hat{Y}_{12}(-\pi). \tag{8}$$

As another example of NMR implementation of single-qubit gate, we analyze the Hadamard gate, which transforms the computational basis states into the equally weighted superposition states. This truly quantum gate can be obtained by the following rotations of a real single qubit:

$$\hat{H} = i\hat{X}(\pi)\hat{Y}(\pi/2) = i\hat{Y}(\pi/2)\hat{Z}(\pi). \tag{9}$$

If the Hadamard gate is applied to the logical qubit $A$ in a quartit, it is described by

$$\hat{H}^A = \hat{H} \otimes \hat{I} = \frac{1}{\sqrt{2}}\begin{bmatrix} 1 & 0 & 1 & 0 \\ 0 & 1 & 0 & 1 \\ 1 & 0 & -1 & 0 \\ 0 & 1 & 0 & -1 \end{bmatrix}, \tag{10}$$

which transforms a general input state $|\psi\rangle$, given by (2), into $\hat{H}^A|\psi\rangle = c_0|+,0\rangle + c_1|+,1\rangle + c_2|-,0\rangle + c_3|-,1\rangle$, where $|\pm\rangle \equiv \frac{1}{\sqrt{2}}(|0\rangle \pm |1\rangle)$. This gate can be realized by the pulse sequence:

$$\hat{H}^A(\theta) = i\hat{Y}_{12}(\pi)\hat{X}_{01}(\pi)\hat{Y}_{01}(\pi/2)\hat{X}_{23}(-\pi)\hat{Y}_{23}(-\pi/2)\hat{Y}_{12}(-\pi) \tag{11}$$

or, equivalently, by

$$\hat{H}^A(\theta) = i\hat{Y}_{12}(\pi)\hat{Y}_{01}(\pi/2)\hat{Z}_{01}(\pi)\hat{Y}_{23}(-\pi/2)\hat{Z}_{23}(\pi)\hat{Y}_{12}(-\pi) \tag{12}$$

If the Hadamard gate is applied to the logical qubit $B$ in a quartit, the transformation is described by

$$\hat{H}^B = \hat{I} \otimes \hat{H} = \frac{1}{\sqrt{2}}\begin{bmatrix} 1 & 1 & 0 & 0 \\ 1 & -1 & 0 & 0 \\ 0 & 0 & 1 & 1 \\ 0 & 0 & 1 & -1 \end{bmatrix}, \tag{13}$$

The action of $\hat{H}^B$ on $|\psi\rangle$ is readily found in analogy to $\hat{H}^A|\psi\rangle$. Hadamard gate $\hat{H}^B$ can be realized by pulses

$$\hat{H}^B(\theta) = i\hat{X}_{01}(\pi)\hat{Y}_{01}(\pi/2)\hat{X}_{23}(\pi)\hat{Y}_{23}(\pi/2), \tag{14}$$

or, equivalently, by
$$\hat{H}^B(\theta) = i\hat{Y}_{01}(\pi/2)\hat{Z}_{01}(\pi)\hat{Y}_{23}(\pi/2)\hat{Z}_{23}(\pi). \quad (15)$$
From the above discussion, it is clear that we will be able to realize any qubit rotation using our NMR devices. However, the principal discussions above ignore the free evolution of the coherent system during the finite pulse duration and the time lag between the pulses. We should compensate for these free evolutions in the pulse sequences for future experimental demonstrations. As already discussed, the pulses can be applied simultaneously if the corresponding operators commute. These features may be helpful in implementing quantum operations based on semiconductor NMR devices.

*3.3. Two logical qubit gates in a quartit*

In addition to single qubit rotations, it is important to describe NMR implementations of CNOT gates for logical qubits in a quartit. The CNOT (XOR) gates are defined as

$$\hat{U}_{CNOT1} = \begin{bmatrix} 1 & 0 & 0 & 0 \\ 0 & 1 & 0 & 0 \\ 0 & 0 & 0 & 1 \\ 0 & 0 & 1 & 0 \end{bmatrix}, \hat{U}_{CNOT2} = \begin{bmatrix} 1 & 0 & 0 & 0 \\ 0 & 0 & 0 & 1 \\ 0 & 0 & 1 & 0 \\ 0 & 1 & 0 & 0 \end{bmatrix} \quad (16)$$

with the control qubit $A$ ($B$) and target qubit $B$ ($A$) for $\hat{U}_{CNOT1}$ ($\hat{U}_{CNOT2}$). The CNOT1 gate transforms the initial/input state $|\psi\rangle$ into $c_0|0\rangle + c_1|1\rangle + c_3|2\rangle + c_2|3\rangle$, while CNOT2 gate changes $|\psi\rangle$ into $c_0|0\rangle + c_3|1\rangle + c_2|2\rangle + c_1|3\rangle$. The following sequences of pulses implement CNOT-like (the so-called Pound-Overhauser CNOT) gates:
$$\hat{U}'_{CNOT1} = \hat{Y}_{23}(\pi), \hat{U}'_{CNOT2} = \hat{Y}_{13}(\pi), \quad (17)$$
which actually differ from standard CNOT gates by the extra minus in one of the off-diagonal terms in (16), which can be corrected (if necessary) by NOT and controlled Z gates. Alternatively, CNOT-like operations can also be realized by
$$\hat{U}''_{CNOT1} = \hat{X}_{23}(\pi), \hat{U}''_{CNOT2} = \hat{X}_{13}(\pi). \quad (18)$$
As shown, it is easier to realize $\hat{U}_{CNOT}$ gates rather than single logical-qubit rotations in a quartit system, which is opposite to the gates for the real two qubit systems. Thus, by virtue of the theorem of Barenco et al. [45], it is seen that arbitrary two (logical) qubit quantum algorithms can in principle be realized in a spin-3/2 NMR system by combining the described single logical qubit rotations and CNOT gates. Note that some other gates can be realized more simply than the described universal gates in our NMR scheme. As an example, let us analyze the SWAP gate, which transforms the initial state $|\psi\rangle$ into $c_0|0\rangle + c_2|1\rangle + c_1|2\rangle + c_3|3\rangle$. The SWAP gate can be obtained by combing CNOT gates as follows:
$$\hat{U}_{SWAP} = \hat{U}_{CNOT1}\hat{U}_{CNOT2}\hat{U}_{CNOT1}. \quad (19)$$
Nevertheless, SWAP-like operations can be realized just by a single pulse as follows:
$$\hat{U}'_{SWAP} = \hat{Y}_{12}(\pi), \text{ or } \hat{U}''_{SWAP} = \hat{X}_{12}(\pi). \quad (20)$$
This differs from the exact SWAP only by the extra minus in one of the off-diagonal terms, which can again be corrected by NOT and controlled Z gates.

Finally, we note that if a system has no quadrupole moment then the three transitions degenerate each other. Such degeneracy would disable NMR quantum information processing within the scheme discussed above.

*3.4. Quantum state tomography based on $M_z$-detection*

Readout is particularly simple if the system is in one of the effective pure states $|k\rangle$ ($k = 0,1,2,3$). However, in general, it is necessary to apply a method known as the quantum state tomography (QST), which enables a complete reconstruction of a given density matrix $\hat{\rho}$ after quantum operations. To reconstruct a density matrix $\hat{\rho}$ for a quartit or two qubits, we need to determine 15 real parameters (The 16th element can be found from the normalization condition.). However, a single NMR readout enables determination of some elements of a given density matrix $\hat{\rho}$ only. We have applied a novel QST scheme based on $M_z$-detection of a spin-3/2 system, which enables the determination of only diagonal elements since $M_z \propto tr(\hat{\rho}\hat{I}_z)$, where $\hat{I}_z = \mathrm{diag}([3/2, 1/2, -1/2, -3/2])$ is the total angular momentum operator for $I = 3/2$. The remaining matrix elements can be obtained by rotating the original density matrix $\hat{\rho}$ through properly chosen rotational operations $\hat{R}_k$ which change $\hat{\rho}$ as follows: $\hat{\rho}^{(k)} = \hat{R}_k \hat{\rho} \hat{R}_k^+$. These operations are performed before NMR readout measurements. Thus, a given density matrix can be reconstructed by transforming $\hat{\rho}$ through various rotations $\hat{R}_k$ in such a way that all elements go over into measurable ones in a given detection scheme.

Quantum state tomography [46, 47] and quantum process tomography [48, 49] were first developed for spin-1/2 systems and later applied to spin-3/2 systems [35, 37, 50]. It should be stressed that, to our knowledge, all known QST methods are based on $M_{xy}$ quadrature detection. As already explained, we directly detect $M_z$ in our system, as the resistance value is proportional to $M_z$. This is completely different from conventional NMR experiments, which detect rotation in XY plane $M_{xy}$.

Here, we consider how to readout the results from an NMR spectrum like that in figure 5(a). The NMR spectrum in this experiment was defined as $\Delta R_{xx}$ measured as a function of r.f. frequency at the pulse duration of $t_{p\pi}$, which approximately corresponds to the π-pulse of the single quantum coherent oscillations. This is equivalent to measuring the NMR spectrum at $t_p \approx 0.06$ ms in figure 3(b). If we can neglect dephasing (usually $t_{p\pi} \ll T_2$ is satisfied) and a small difference in oscillation frequency between $\Omega_{R1}(=\Omega_{R3})$ and $\Omega_{R2}$ [41], three peaks in the NMR spectrum approximately represent the subtracted population densities, i.e., $\rho_{11} - \rho_{00}$, $\rho_{22} - \rho_{11}$ and $\rho_{33} - \rho_{22}$, where $\rho_{mn} \equiv \langle m|\hat{\rho}|n\rangle$ and $\rho_{ii}$ correspond to the density of $i$ spin states and $i=0, 1, 2$ and $3$ corresponds to $|00\rangle$, $|01\rangle$, $|10\rangle$ and $|11\rangle$ of the equivalent two-qubit system.

Taking into account this readout scheme, we propose the use of the following natural set of 12 rotations for the $M_z$-based tomography of spins-3/2:

$$\hat{X}_{01}(\pi/2), \hat{Y}_{01}(\pi/2), \hat{X}_{12}(\pi/2), \hat{Y}_{12}(\pi/2), \hat{X}_{23}(\pi/2), \hat{Y}_{23}(\pi/2),$$
$$\hat{X}_{02}(\pi/2), \hat{Y}_{02}(\pi/2), \hat{X}_{13}(\pi/2), \hat{Y}_{13}(\pi/2), \hat{X}_{03}(\pi/2), \hat{Y}_{03}(\pi/2). \qquad (21)$$

The choice of these rotations can be understood as follows: When we apply a $\hat{Y}_{01}(\pi/2)$ pulse after a certain quantum operation, we will get a diagonal component containing $\rho_{01}$ and $\rho_{10}$. In our NMR spectrum based on $M_z$ detection, one of the three signals is proportional to $\mathrm{Re}(\rho_{01}) + \mathrm{Re}(\rho_{10})$. Because $\rho_{ij} = \rho_{ji}^*$, we can get $\mathrm{Re}(\rho_{01}) = \mathrm{Re}(\rho_{10})$. Similarly, $\hat{Y}_{12}(\pi/2)$, $\hat{Y}_{23}(\pi/2)$, $\hat{Y}_{02}(\pi/2)$, $\hat{Y}_{13}(\pi/2)$ and $\hat{Y}_{03}(\pi/2)$ give us $\mathrm{Re}(\rho_{12}) = \mathrm{Re}(\rho_{21})$, $\mathrm{Re}(\rho_{23}) = \mathrm{Re}(\rho_{32})$, $\mathrm{Re}(\rho_{02}) = \mathrm{Re}(\rho_{20})$, $\mathrm{Re}(\rho_{13}) = \mathrm{Re}(\rho_{31})$ and $\mathrm{Re}(\rho_{03}) = \mathrm{Re}(\rho_{30})$. Imaginary parts are also estimated by applying a $\hat{X}_{ij}(\pi/2)$ pulse and using $\mathrm{Im}(\rho_{ij}) = -\mathrm{Im}(\rho_{ji})$. Although the method requires 12 measurements as well as two- and three-photon operations, the scheme seems simple and easy to understand. This method can essentially be optimized by reducing the number of required rotations to six while increasing the reconstruction sensitivity as well as its relative error. The details of this novel tomography scheme are beyond the scope of this paper and will be discussed elsewhere [51]. Moreover, the three- and two-photon operations can be replaced by solely single-photon operations, as we show in the Appendix.

**4. Preparation of effective pure states**

The initialization of an NMR quantum computer is based on the preparation of effective pure (psuedopure) states for the nuclei ensemble. In the effective pure states, the deviation density matrix can be treated as if it is in a pure state and any unitary evolution, i.e. quantum computation, can be applied to that pure state.

In conventional NMR quantum information experiments, the effective pure states are prepared from thermal equilibrium states. A review of these methods applied to spin-1/2 systems can be found in [14]. A generalized method for spin-3/2 liquid-state systems was first used experimentally in [29]. Three approaches have been developed for preparation of effective pure states: logical labeling using ancillary qubits as labels [11], spatial averaging [52], and temporal averaging [53]. However, the effective pure states starting from the thermal equilibrium situation have the available pure state population of only $N\hbar\omega_0/(4k_BT)$, where $k_B$ is Boltzman's constant, $T$ the temperature, and $N$ the total number of nuclear spins. The $\hbar\omega_0$ is on the order of 0.1 μeV and usually much smaller than $4k_BT$.

The situation is different in our NMR semiconductor devices. We can use dynamic nuclear spin polarization induced by the current flow and prepare the effective pure states from this strongly polarized situation, which makes a large population available as the pure state. Here, we used the $^{69}$Ga NMR spectrum to estimate the population among four spin states. As already discussed, the NMR spectrum in this experiment was defined as $\Delta R_{xx}$ measured as a function of radio frequency at the pulse duration of $t_{p\pi}$, which approximately corresponds to the π-pulse of the single quantum coherent oscillations. The amplitudes of three NMR peaks roughly represent the subtracted density distributions, i.e., $\rho_{11} - \rho_{00}$, $\rho_{22} - \rho_{11}$ and $\rho_{33} - \rho_{22}$. Figure 5(a) shows the NMR spectrum, i.e. the subtracted density distribution, just after dynamical polarization by the current flow. As schematically shown in the inset, the population becomes larger for the higher spin states, reflecting dynamical polarization. However, this situation is far from the effective pure states. We therefore describe how to create the effective pure states by applying appropriate pulse sequences.

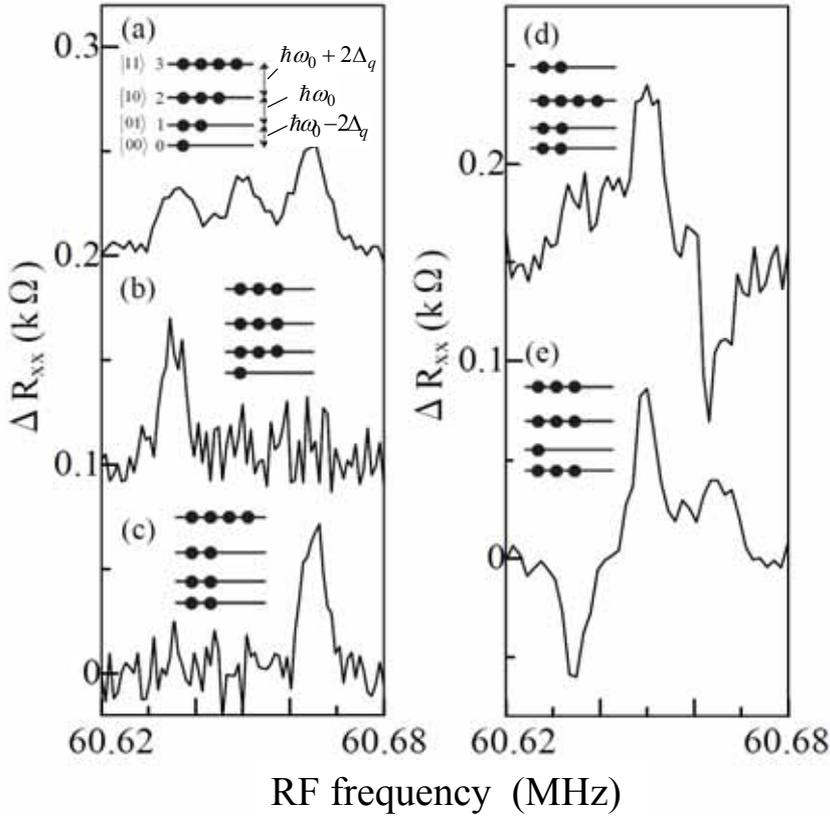

**Figure 5** Pulsed NMR spectra of $^{69}$Ga corresponding to the initial state (a) and four effective pure states (b)-(e). The spectrum was obtained by applying pulses with 120-μsec duration after each initialization process. The population of the nuclear spins in the four levels is schematically shown in the insets. All spectra were taken at 6 T and 150 mK, and the observed quadrupolar splitting was 15 kHz.

Although the degeneracy of electron spin states is very important for inducing nuclear spin and electron spin interaction as already discussed, the interaction occurs in a rather wide range around the degeneracy point [40] and how the population changes between four spin states strongly depends on the conditions for the dynamic polarization [54]. Here, we selected the condition where the population increases linearly with spin states. In other words, the amplitude is almost constant among the three NMR peaks like in the spectrum shown in figure 5(a). Starting from such an initial state, we can create effective pure states for $|00\rangle$, $|01\rangle$, $|10\rangle$, and $|11\rangle$ as shown in figures 5(b) to (e), using frequency selective pulse sequences. In preparing the $|00\rangle$ ($-|00\rangle$ in the strict definition) effective pure state for example, we apply a pulse with frequency of $\omega_0$ [we denote as $\hat{X}_{12}(\pi)$] to interchange the population of levels 1 and 2 and apply a $\pi/2$ pulse with frequency of $\omega_0 + 2\Delta_q/\hbar$ [$\hat{X}_{23}(\pi/2)$] to equalize the populations of levels 2 and 3. If we allow the two-photon transition, $-|00\rangle$ state is prepared by applying a $\hat{X}_{13}(\pi/2)$ single pulse. Similarly, the effective pure state $|11\rangle$ is obtained by sequentially applying $\hat{X}_{12}(\pi)$ and $\hat{X}_{01}(\pi/2)$ pulses or applying a $\hat{X}_{02}(\pi/2)$ single pulse. For $|10\rangle$ and $|01\rangle$ states, we use pulse sequences containing two-quantum transitions $\hat{X}_{23}(\pi) \hat{X}_{02}(\pi/2)$ and $\hat{X}_{01}(\pi) \hat{X}_{13}(\pi/2)$, respectively. We can also generate the effective $|01\rangle$, $|10\rangle$, and $|00\rangle$ pure states from the effective $|11\rangle$ pure state. Namely, after generating $|11\rangle$ state, the other $|ij\rangle$ state can be obtained by applying the pulse $\hat{X}_{2i+j,3}(\pi)$ which simply switches the populations between the states $|11\rangle$ and $|ij\rangle$.

Here, we have shown the simple scheme for forming an effective pure states from the linearly distributed population. However, it is clear that we can form the effective pure state from any arbitrary distribution by applying more complicated pulse sequences. Although any pulse manipulates nuclear spin states coherently, the designed population is maintained up to $T_1$ longer than 100 seconds [25]. Therefore, it is possible to form effective pure states without complicated coherent effects by setting appropriate waiting time $T_2 < t < T_1$ during the pulse sequences.

## 5. Conclusions

In our pursuit of all-electrical control and detection of nuclear spins in semiconductors, we have found that the fractional-quantum-Hall regime around $\nu=2/3$ can be used to control and detect nuclear spin states by electron-nuclear spin coupling. The degenerate condition between spin polarized and unpolarized $\nu=2/3$ states results in pronounced hyperfine interactions. We have extended these interactions to a nanoscale device, in which a point-contact is formed by using split gates and an antenna gate is integrated for applying electromagnetic radiation to the point-contact region. By passing an electrical current through the structure, nuclear spins can be selectively polarized in the point-contact region, where current density is high. Furthermore, the resistance of the point contact shows changes that are approximately proportional to the vertical magnetization originating from the nuclear spins. Accordingly, if an alternating current is driven through the antenna gate to apply electromagnetic radiation at the NMR frequency, coherent nuclear spin oscillations occur only at the desired transition, resulting in oscillations of the nuclear spin magnetization, which in turn is detected by the resistance of the point contact. Strikingly clear coherent oscillations are observed, reflecting all possible transitions among the four nuclear spin states of each nuclide (namely, $^{69}$Ga, $^{71}$Ga and $^{75}$As). The arbitrary control of superposition among four spin levels equals two-qubit operation. We have theoretically shown how to make the two fictitious qubits from a four-level system and how to design the operation pulses for one- and two-qubit operations. A state tomography scheme is also proposed for the novel $M_z$ detection realized in our device. As the starting point of the quantum operation, we have demonstrated the formation of the effective pure states experimentally. Decoherence is the most important issue for solid state quantum systems, and we have estimated the decoherence from the Rabi-type oscillation of our NMR device. We found that As-Ga direct dipole coupling is a less important factor in determining the decoherence due to the magic angle of the crystal bonds. The $T_2$ time is extended to values longer than 1.5 ms by the electron decoupling. Further improvements of $T_2$ will be allowed by using more sophisticated decoupling techniques like the one in [55]. The discussions in this paper clearly suggest that the exciting features of nanometre-scale NMR devices make them plausible candidates for the future quantum information processing based on solid-state systems.

**Acknowledgements**
The authors are grateful to K. Takashina, K. Hashimoto, T. Saku, S. Miyashita and N. Kumada for fruitful collaborations. YH is partially supported by a Grant-in-Aid for Scientific Research from JSPS. AM is supported by the Polish State Committee for Scientific Research under grant No. 1 P03B 064 28.

**Appendix: Quantum operations without two- and three-photon transitions**

Here, we suggest a way to replace three- and two-photon operations by single-photon ones, which, in particular, can solve the problem related to the degeneracy of single- and three-photon transitions at $\omega_0$.

As already explained, we can observe (i) two-photon (i.e., double quantum) transitions between levels $|0\rangle$ and $|2\rangle$ at frequency $\omega_0 - \Delta_q/\hbar$ and between levels $|1\rangle$ and $|3\rangle$ at frequency $\omega_0 + \Delta_q/\hbar$, as well as a (ii) three-photon (triple quantum) transition at frequency $\omega_0$ between levels $|0\rangle$ and $|3\rangle$. In the latter case, there is a problem of degeneracy with the transition between $|1\rangle$ and $|2\rangle$, which implies that the pulse at frequency $\omega_0$ causes the transition not only between levels $|0\rangle$ and $|3\rangle$ but also an undesirable transition between levels $|1\rangle$ and $|2\rangle$. It is possible to select the oscillating field strength $B_1$ that satisfies some angle rotation for the transition between $|0\rangle$ and $|3\rangle$ but multiple of $2\pi$ for the transition between $|1\rangle$ and $|2\rangle$. Therefore, it is possible to realize pure $|0\rangle$-$|3\rangle$ operation without $|1\rangle$-$|2\rangle$ rotation. However, the current amplitude necessary for this operation may be high and the operation is not realistic from the viewpoint of heating. The simplest way to avoid such degeneracy is just to replace rotations requiring three-photon transitions by combinations of rotations based only on (i) single- and two-photon transitions or (ii) just single-photon transitions. We find that the following replacements can be used:

$$\hat{X}_{03}(\theta) = \hat{Y}_{13}(\pi)\hat{X}_{01}(\theta)\hat{Y}_{13}(-\pi) = \hat{Y}_{02}(\pi)\hat{X}_{23}(-\theta)\hat{Y}_{02}(-\pi)$$
$$= \hat{Y}_{01}(\pi)\hat{X}_{13}(-\theta)\hat{Y}_{01}(-\pi) = \hat{Y}_{23}(\pi)\hat{X}_{02}(\theta)\hat{Y}_{23}(-\pi)$$
$$= \hat{Y}_{01}(\pi)\hat{Y}_{23}(\pi)\hat{X}_{12}(-\theta)\hat{Y}_{23}(-\pi)\hat{Y}_{01}(-\pi) = \cdots,$$

where $\theta$ stands for arbitrary angle. Thus, we do not need to use any strong pulses to make transition $|0\rangle$-$|3\rangle$, which can erroneously induce transition $|1\rangle$-$|2\rangle$. Moreover, if one prefers to use only rotations based on single-photon transitions rather than both single- and two-photon transitions, then we can replace the latter as follows:

$$\hat{X}_{13}(\theta) = \hat{Y}_{23}(\pi)\hat{X}_{12}(\theta)\hat{Y}_{23}(-\pi) = \hat{Y}_{12}(\pi)\hat{X}_{23}(-\theta)\hat{Y}_{12}(-\pi),$$
$$\hat{Y}_{13}(\theta) = \hat{Y}_{23}(\pi)\hat{Y}_{12}(\theta)\hat{Y}_{23}(-\pi) = \hat{Y}_{12}(\pi)\hat{Y}_{23}(-\theta)\hat{Y}_{12}(-\pi),$$
$$\hat{X}_{02}(\theta) = \hat{Y}_{12}(\pi)\hat{X}_{01}(\theta)\hat{Y}_{12}(-\pi) = \hat{Y}_{01}(\pi)\hat{X}_{12}(-\theta)\hat{Y}_{01}(-\pi),$$
$$\hat{Y}_{02}(\theta) = \hat{Y}_{12}(\pi)\hat{Y}_{01}(\theta)\hat{Y}_{12}(-\pi) = \hat{Y}_{01}(\pi)\hat{Y}_{12}(-\theta)\hat{Y}_{01}(-\pi).$$

One may raise an objection about a reverse process, namely, whether it is possible to induce selectively the transition $|1\rangle$-$|2\rangle$ without causing the transition $|0\rangle$-$|3\rangle$. The rotation frequency is proportional to the 1st Bessel function of the oscillation field $B_1$ for the coherent rotation between $|1\rangle$ and $|2\rangle$. In contrast, it is proportional to the 3rd Bessel function for that between $|0\rangle$ and $|3\rangle$. Therefore, the transition $|0\rangle$-$|3\rangle$ becomes negligible if the applied pulse (at frequency $\omega_0$) is weak, although this scheme is limited by the finite $T_2$ in our system.